\documentclass[onecolumn,authoryear]{els-mrw} 

\usepackage{amsmath,amssymb,amsfonts,amsthm,makeidx,graphicx}
\usepackage{txfonts}
\usepackage{helvet}
\usepackage{subcaption}
\usepackage{tablefootnote}
\newcommand{\nat}{Nature}

\newcommand{\apj}{Astrophys.~J.}      
\newcommand{\aj}{Astron.~J.}      
\newcommand{\aaps}{Astron.~Astrophys.~Supp.}  
\newcommand{\aap}{Astron.~Astrophys.}  
\newcommand{\aapr}{Astron.~Astrophys. Rev.}  
\newcommand{\apjl}{Astrophys.~J.~Lett}   
\newcommand{\aplett}{Astrophys.~Lett} 
\newcommand{\apjs}{Astrophys.~J.~Supp.}   
\newcommand{\mnras}{Mon.~Not.~R.~Astron.~Soc.} 
\newcommand{\actaa}{Acta Astronomica} 

\newcommand{\pasp}{Publ.~Astron.~Soc.~Pac.}   
\newcommand{\araa}{Annu.~Rev.~Astron.~Astrophys} 
\newcommand{\qjras}{Q.~J.~R.~Astron.~Soc.} 

\newcommand{\zap}{Zeitschrift f{\"u}r Astrophysik}
\newcommand{\planss}{Planetary~\&~Sp.~Sci.} 

\newcommand\kms{{\rm\,km\,s^{-1}}}

 \def\mso{\,M_\odot}

\begin{document}

\chapter{Blue straggler stars}\label{chap1}

\author[1]{Chen Wang}%
\author[1,2,3]{Taeho Ryu}%

\address[1]{\orgname{Max Planck Institute for Astrophysics}, \orgaddress{Karl-Schwarzschild-Strasse 1, 85748 Garching, Germany}}
\address[2]{\orgname{JILA, University of Colorado and National Institute of Standards and Technology}, \orgaddress{440 UCB, Boulder, 80308, CO, USA}}
\address[3]{\orgname{Department of Astrophysical and Planetary Sciences}, \orgaddress{391 UCB, Boulder, 80309, CO, USA}}

\articletag{Chapter Article tagline: update of previous edition,, reprint..}

\maketitle

\begin{glossary}[Glossary]
\term{Main-sequence star} Stars that are burning hydrogen in their cores. 

\term{Globular cluster} Globular clusters are densely packed, spherical collections of thousands to millions of old low-mass stars, which are mostly found in the galactic halo.  

\term{Open cluster} Open clusters are loosely packed groups of hundreds to thousands of relatively young massive stars, which are mostly found in the galactic plane. 

\term{Cluster main-sequence turn off} The point on the Hertzsprung-Russell diagram where stars in a star cluster cease to follow the main sequence and begin to evolve into later stages.

\term{Rejuvenation} Rejuvenation refers to the process by which a star appears younger than its actual age due to certain events or mechanisms that modify its evolutionary state. 

\end{glossary}

\begin{glossary}[Nomenclature]
\begin{tabular}{@{}lp{34pc}@{}}
BSS & Blue straggler star\\
GC & Globular Cluster\\
OC & Open Cluster\\
MS &Main-sequence \\
HRD & Hertzsprung-Russell diagram\\
CMD & Color-magnitude diagram\\
AGB & Asymptotic giant branch\\
MHD & Magnetohydrodynamic\\
SPH & Smoothed particle hydrodynamic\\
W UMa star & W Ursae Majoris variable\\
\end{tabular}
\end{glossary}

\begin{abstract}[Abstract]
Blue straggler stars are unique main-sequence stars that appear more luminous, hotter, and therefore younger, than their coeval counterparts. In star clusters, these stars are located above the cluster turn-off in the Hertzsprung-Russell diagram or color-magnitude diagram. First identified in the 1950s, these stars are found across diverse environments, from sparse galactic fields to dense star clusters. They are crucial for understanding stellar and binary evolution and star cluster dynamics. Despite extensive research, many challenges concerning their properties and origin mechanisms remain unresolved. This chapter delves into the properties and origins of blue stragglers, examining how theoretical tools are employed to study them and the implications of each proposed formation mechanism. We assess how contemporary observational data either support or challenge these theoretical predictions. Continued theoretical and observational efforts are essential for advancing our understanding of these enigmatic stars.

\end{abstract}
\begin{BoxTypeA}[chap1:box1]{Key Points}
\begin{itemize}
\item Blue straggler stars appear younger than their contemporaries, indicating that rejuvenation processes are essential for their formation.
\item The primary mechanisms for blue straggler star formation include binary mass transfer, binary mergers and stellar collisions, each contributing distinct properties to the resulting stars.
\item The properties of blue straggler stars formed through various mechanisms have been explored using a combination of 3D and 1D simulations, as well as N-body simulations.
\item Current observations reveal that all these mechanisms play a role, with their significance varying across different environments. 
\item The discrepancies between theoretical predictions and observational data highlight a significant demand for further studies on both the theoretical and observational sides.
\end{itemize}
\end{BoxTypeA}

\section{Introduction}
\begin{figure}[t]
\centering
\includegraphics[width=.9\textwidth]{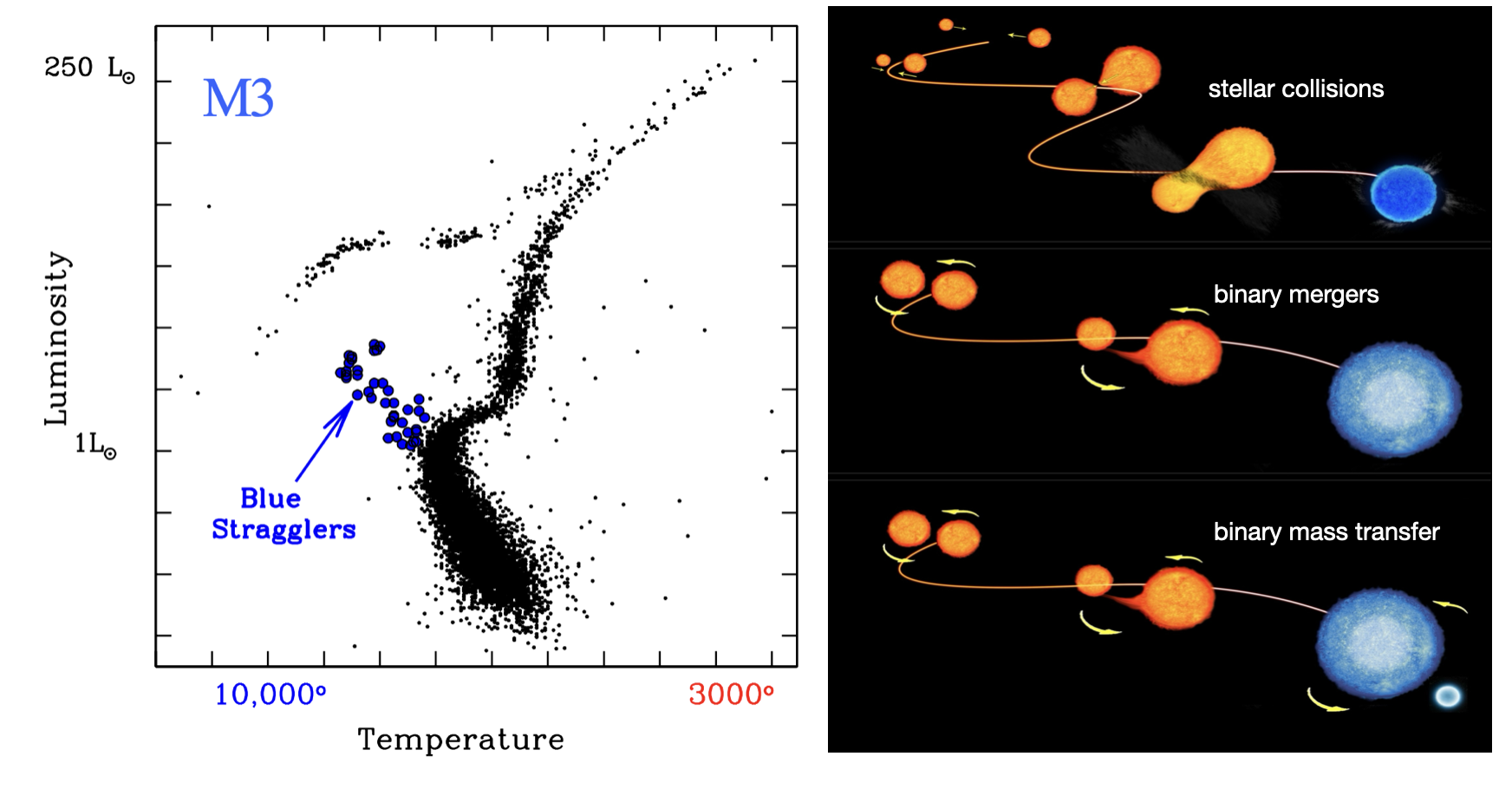}
\caption{Left: Distribution of blue straggler stars in the globular cluster M\,3 in the Hertzsprung-Russell diagram. The blue and black dots correspond to the blue straggler stars and other stars in this cluster, respectively. Right: Artistic illustration of the formation mechanisms for blue straggler stars: stellar collisions (top), binary mergers (middle) and binary mass transfer (bottom). This figure is adapted from Fig.\,1 in \cite{2020RLSFN..31...19F}.}
\label{fig:BSS}
\end{figure}

In the classical definition, blue stragglers stars (BSSs) are a fascinating class of stellar anomalies observed in star clusters, positioned above the cluster main-sequence (MS) turn-off in the Hertzsprung-Russell diagram (HRD) or color-magnitude diagram (CMD) (see Fig.\,\ref{fig:BSS}, left panel). In our traditional view, all stars in a star cluster are believed to be born from a single starburst and therefore have the same age. Their positions in the HRD/CMD indicate that these stars are brighter, bluer, and appear younger compared to their contemporaries. BSSs were first identified in the old globular cluster (GC) M\,3 in 1953 \citep{1953AJ.....58...61S}. The term `straggler' was first coined by \cite{1958ApJ...128..174B} to describe their surprisingly youthful appearance relative to other cluster stars. 
Although initially observed in clusters, the concept of BSSs has since been extended to include similarly anomalous field stars, though identifying these field BSSs is more challenging due to the need for other methods to measure their real age, for instance, using the age of comparable metallicity GCs. 

To date, extensive photometric data have revealed large populations of BSSs across various stellar environments, from sparse galactic fields, open clusters (OCs) to densely populated GCs, encompassing both high-mass and low-mass stars. The most comprehensive data sets exist for GCs. A catalogue of about 3000 BSSs in 56 GCs can be found in \cite{2004ApJ...604L.109P}. However, more in-depth studies on properties of BSSs, such as binary fraction and stellar rotation, have been conducted in OCs, thanks to their proximity which facilitates high-quality spectroscopic observations. 
Using photometric data, more than 1600 BSS candidates have been identified in nearly 400 OCs \citep{2007A&A...463..789A,2021A&A...650A..67R,2021MNRAS.507.1699J,2023A&A...672A..81L}. Among them, NGC\,188 ($\sim$ 7\,Gyr) and M\,67 ($\sim$ 4\,Gyr) serve as prime examples where detailed spectroscopic studies of stellar properties have been undertaken \citep{2009Natur.462.1032M,2008AJ....135.2264G,1984ApJ...279..237P,2007HiA....14..444L}. 
Field BSSs are predominantly found in metal-poor populations, such as the Galactic halo. This is because the ages of metal-poor stars are closely related to their metallicities, making it easier to identify BSSs by comparing them with GCs that have similar metallicities. In contrast, metal-rich stars are more age-heterogeneous. The Sloan Digital Sky Survey (SDSS) has detected thousands of BSSs throughout the Galactic halo \citep{2000ApJ...540..825Y}.

In general, the position of a star in a star cluster in the HRD/CMD correlates with its mass, with more massive stars being more luminous and hotter, although the exact location depends on the assumptions on stellar evolution. BSSs, which appear more massive than their normal MS counterparts within the same cluster, are thus thought to form through processes that add mass to a normal MS star. Potential mechanisms include accretion during binary mass transfer \citep{1964MNRAS.128..147M,2007AIPC..948..431C,2008MNRAS.387.1416C,2015ASSL..413..179I,2022A&A...659A..98S}, mergers from binary interactions, i.e. interactions between the two stars \citep{1964MNRAS.128..147M,2001ApJ...552..664N,2016MNRAS.457.2355S,2019Natur.574..211S}, or mergers from stellar collisions \citep{1976ApL....17...87H,1987ApJ...323..614B,1989AJ.....98..217L,1997ApJ...487..290S,2005MNRAS.358..716S}, as illustrated in the right panel of Fig.\,\ref{fig:BSS}. It is believed that all these mechanisms contribute to the formation of BSSs, but their importance varies based on many factors, including the stellar density of environments. In particular, stellar collisions are more common in dense environments such as GCs and cores of OCs, whereas binary mass transfer and binary evolutionary-induced mergers (hereafter binary mergers) dominate in less crowded environments, like galactic fields and the outskirts of OCs. In some clusters, such as GC M\,30, a combination of mechanisms, including both stellar collisions and binary mass transfer, is necessary to account for the observed BSSs \citep{2009Natur.462.1028F}. 

The study of BSSs intersects various domains of astronomical research. Theoretically, an array of simulation tools, including 3D, 1D, and N-body simulations, has been utilized to investigate their origins \citep{1987ApJ...323..614B,1996ApJ...468..797L,1997ApJ...487..290S,2005MNRAS.358..716S,2005MNRAS.363..293H,2019Natur.574..211S}. 3D simulations play a crucial role in exploring the properties of merger products, as merger is a dynamical process. These simulations demonstrate that merger products may exhibit strong magnetic fields and rapid rotation immediately following the merger \citep{2019Natur.574..211S}. However, the subsequent evolution of the magnetic fields and stellar spins are not well understood. Results from 3D simulations are often integrated into 1D stellar evolution codes to simulate the further evolution of these merger products. Such studies show that BSSs formed from mergers do not retain clear `memories' of their origins, and have similar properties as genuine single stars of equivalent masses \citep{2013MNRAS.434.3497G}. Conversely, BSSs from binary mass transfer, modeled using 1D simulations, are expected to show distinct properties, such as rapid rotation, unusual surface chemical compositions, and the presence of companion stars \citep{2001ApJ...552..664N,2004MNRAS.355.1182C}. While 3D and 1D simulations are used for studying isolated systems, N-body simulations complement these studies by addressing the dynamical evolution of star clusters, which are crucial for our understanding of BSSs in dense GCs \citep{2005MNRAS.358..572I,2005MNRAS.363..293H}.

From the observational point of view, high-quality photometric and spectroscopic data are essential for accurately determining the properties of BSSs and discerning their formation mechanisms. Key observables include their frequencies across various environments, mass estimates, spatial distributions within star clusters, binary or higher multiplicity status, rotational velocities, and chemical compositions. Typically, a comprehensive understanding of their origins requires integrating multiple observational properties. 

In this chapter, we first summarize the current observations of BSSs across various environments in Section\,\ref{sec:ob}. In Section\,\ref{sec:form} we discuss theoretical efforts to  investigate the formation mechanisms of BSSs. Section\,\ref{sec:simu} shows current simulations modeling the BSSs in star clusters and their comparison with observational data. Then we briefly describe the later evolution of BSSs after hydrogen exhaustion in Section\,\ref{sec:later}, and summarize the key points of the chapter and offer future perspectives in Section\,\ref{sec:sum}.

\section{Observations of BSSs}\label{sec:ob}
The identification of BSSs primarily relies on photometric observations, which measure the brightness and color of stars. These observations also allow for estimating their frequency relative to other stars, their spatial distributions, and their masses as derived from positions in the CMD compared to theoretical stellar models. However, other properties, including multiplicity, rotational velocity, and chemical composition, are crucial for deciphering the origins of BSSs and understanding the underlying physics. To obtain these properties, high-quality spectroscopic observations are essential.
Although systematic spectroscopic observations are limited, observations of well-studied OCs like NGC\,188 and M\,67 have provided valuable insights. In this section, we offer a concise overview of the current observational data on BSSs and discuss its implications for understanding their origins. The key properties discussed here are summarized in Table\,\ref{tab:1}, which is adapted from Table 11.1 in \cite{2015ASSL..413..251P}.

\begin{table}
\begin{center}
\caption{Summary of BSSs properties in different environments. The table is adapted from Table 11.1 in \cite{2015ASSL..413..251P}.}
\label{tab:1}
\begin{tabular}{ l c c c c c c c} 
\toprule
\midrule
Environment  & Globular cluster & Open cluster & Field \\
\midrule
Frequency\footnotemark[1]  &  $10^{-5}-10^{-4}$ & $10^{-3}-10^{-2}$ &--\\
Spatial distribution  &  Centrally concentrated   & Centrally concentrated & -- \\
    &   and sometimes bi-modal & and sometimes bi-modal &  \\
Binarity & Unknown; Large fraction in eclipsing binaries & High; 79\% in M\,67;76\% in NGC\,188 & High \\
Companion  mass  &  unknown  & Peaks at $0.55\,M_\odot$ & $0.18 - 0.55\,M_\odot$ \\
Period & unknown & $P> 500$ days & typical 200-800 days \\
      &     &   10\% $P<10$ days & many $P<10$ days \\
Eccentricity & unknown & High; $<e>\sim 0.34$ & Low; $<e>\sim 0.17$ \\
\bottomrule
\end{tabular}
\end{center}
\footnotesize{1. Frequency is defined as the ratio between the number of blue stragglers and all stars in a cluster.}
\end{table}

\subsection{Frequency}
The frequency of BSSs is in general relatively low. It varies significantly across different environments, with higher values in lower-density environments. One possible explanation for this variation is that multiple systems, the progenitors of BSSs, are more likely to be disrupted in denser environments \citep{2000AJ....120.1014P}. In OCs, the typical BBS number is a few tens, corresponding to a BSS fraction of a few $10^{-3}$. In contrast, GCs generally host a few to hundreds of BSSs within populations of $10^5 - 10^6$ stars, resulting in a much lower frequency of a few $10^{-5}-10^{-4}$. 
The frequency of field BSSs remains less defined due to the difficulties in identifying these stars. \cite{2000AJ....120.1014P} suggested that field BSSs are more common per unit luminosity than those in the outer regions of GCs. 

Analyzing the relationship between BSS frequency and cluster properties such as cluster mass, binary fraction and collision rate can offer valuable insights into their formation mechanisms. For instance, if BSSs in a cluster core primarily arise from stellar collisions, a positive relation between the frequency of BSSs and collision rate is expected. Alternatively, if BSSs mainly result from binary evolution, their frequency is expected to be proportional to the total cluster mass and binary fraction. 

In OCs, the number of BSSs correlates with the binary fraction and cluster mass, indicating that binary evolution dominates the formation of BSSs in these clusters \citep{2024A&A...685A..33R}. A similar correlation has been observed in low-density GCs \citep{2008A&A...481..701S}. 
Unexpectedly, a comprehensive analysis of BSSs in the cores of 57 GCs also showed no clear correlation between the number of BSSs and collision rates, but a significant positive relation between the number of BSSs and the total stellar mass within cluster cores \citep{2009Natur.457..288K}. This finding underscores the importance of binary evolution in BSS formation, even in dense cluster cores. Nonetheless, these findings do not discount the role of stellar collisions, especially since binary fractions in dense environments are influenced by dynamical encounters. 

\subsection{Mass}
BSSs can be either high-mass or low-mass stars, depending on the environment in which they are found. Theoretically, BSSs are expected to be more massive than the cluster turn-off mass.
Determining stellar masses is crucial yet challenging in astronomy. While binary systems with well-defined orbits offer the most direct and reliable way for measuring stellar masses, identifying binaries and precisely determining their orbital motions are difficult. 

In many cases, the masses of BSSs are inferred by comparing their positions in the CMD with theoretical stellar models. This method naturally leads to the conclusion that BSSs are more massive than the other normal MS stars in the same cluster. For instance, in the OC NGC\,188, CMD-derived masses for BSSs range from 1.15 to 1.55$\,M_\odot$, which are notably larger than the cluster's turn-off mass of approximately 1$\,M_\odot$ \citep{2012AJ....144...54G}.

Spectroscopic observations offer another model-dependent method for determining stellar masses. This technique has been pivotal in studying BSSs, such as in the case of the GC\,47 Tucanae, where spectroscopy revealed a BSS nearly twice the mass of the cluster's turn-off mass \citep{1997ApJ...489L..59S}. 

Furthermore, asteroseismology provides essential information into stellar masses by analyzing the pulsation frequencies of stars. This method has been applied to BSSs in the GC NGC\,6541, where nine SX Phoenicis variables were identified \citep{2014ApJ...783...34F}. Their masses were estimated to be between 1.0 and 1.1$\,M_\odot$, significantly larger than the cluster turn-off mass of around $0.75\,M_\odot$.

\subsection{Radial distribution}
In a star cluster, the radial distribution of BSSs is determined by two factors: their formation places and their later evolution in the cluster. The formation location may depend on their formation mechanisms. Specifically, BSSs resulting from stellar collisions are expected to appear in the cluster center where stellar densities are highest. In contrast, BSSs that arise from binary evolution are expected to form throughout the cluster.
After their formation, BSSs are expected to migrate towards the cluster center due to their higher masses, a process known as mass segregation. Consequently, the radial distribution of BSSs is a valuable indicator of a cluster's dynamical evolution \citep{2012Natur.492..393F}. 

\begin{figure}[t]
\centering
\begin{subfigure}[b]{0.32\textwidth}
\includegraphics[width=1.0\textwidth]{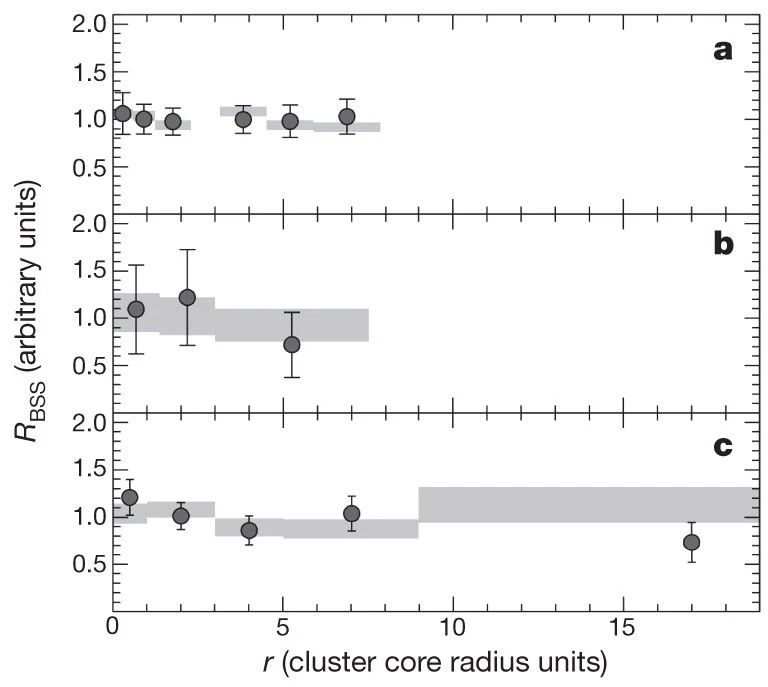}
\end{subfigure}
\begin{subfigure}[b]{0.3\textwidth}
\includegraphics[width=1.0\textwidth]{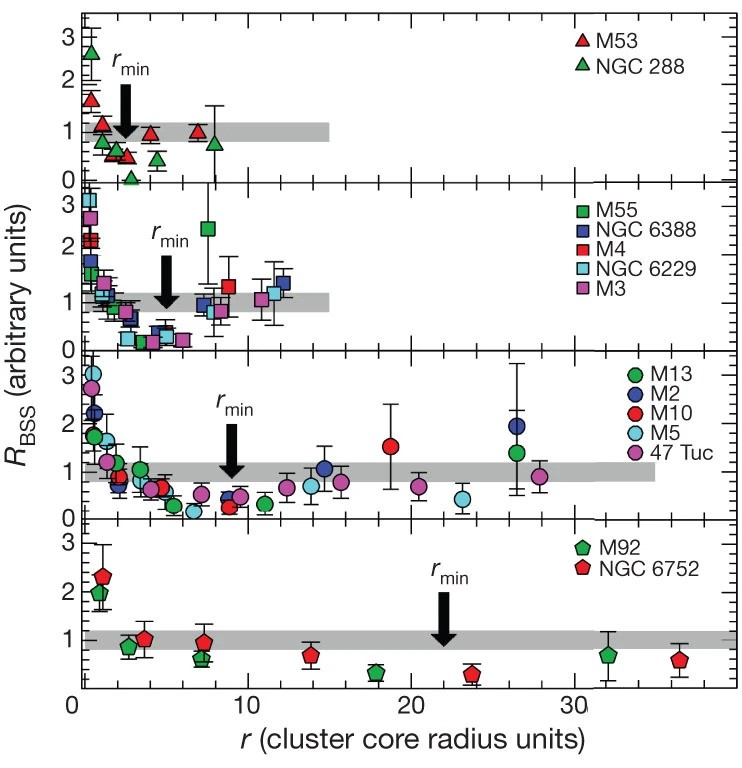}
\end{subfigure}
\begin{subfigure}[b]{0.32\textwidth}
\includegraphics[width=1.0\textwidth]{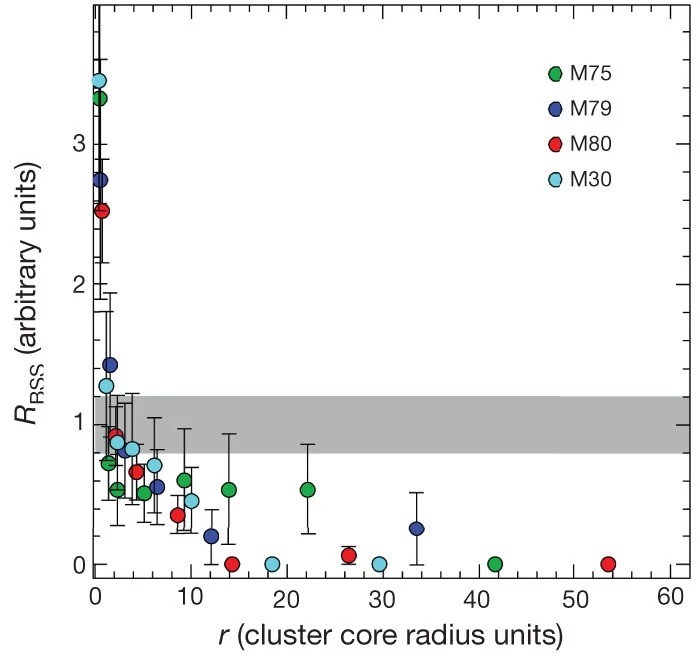}
\end{subfigure}
\caption{Radial distribution of blue straggler stars across clusters with different dynamical ages. The normalized ratio $R_\mathrm{BSS}$ is defined as $R_\mathrm{BSS}(r)=[N_\mathrm{BSS}(r)/N_\mathrm{BSS, tot}][L_\mathrm{samp}(r)/L_\mathrm{samp,tot}]$, where $N_\mathrm{BSS}(r)$ and $N_\mathrm{BSS, tot}$ represent the number of blue straggler stars measured in a specific radial bin and the total number in the cluster, respectively. $L_\mathrm{samp}(r)$ and $L_\mathrm{samp,tot}$ are the luminosity sampled in the same radial bin and the total sampled luminosity, respectively. The grey areas indicate the ratios for reference stars. The panels, from left to right, represent clusters with increasing dynamical ages. This figure is reproduced from Figs.\,1,2 and 3 in \cite{2012Natur.492..393F}.}
\label{fig:rad_dis}
\end{figure}

Figure\,\ref{fig:rad_dis} demonstrates how the spatial distribution of BSSs within their host clusters can reveal the clusters' dynamical ages. A uniform radial distribution of BSSs, similar to that of the general star population (reference stars), suggests that the cluster is dynamically young. As clusters evolve dynamically, the influence of dynamical friction first becomes noticeable near the cluster center and gradually affects more distant regions.
In clusters that are more dynamically evolved, this friction causes BSSs to concentrate near the center, creating a distribution with a central peak and a dip at intermediate radii. The most distant BSSs in these clusters remain unaffected by dynamical friction. Therefore, a bimodal distribution with peaks at both the center and the outer edges is seen.
In highly evolved clusters, dynamical friction impacts even the outermost BSSs, pulling them toward the center and resulting in a distribution that predominantly shows a central peak.


\subsection{Multiplicity}\label{sec:ob_multiplicity}
Determining the binary status of BSSs is essential for understanding their formation mechanisms. BSSs resulting from binary mass transfer usually have close companions, typically compact objects, unless the binary system was disrupted by supernova kicks. The orbits of these systems tend to be circular due to the mass transfer process. On the other hand, BSSs formed through mergers, either from stellar collisions or binary mergers, typically do not have close companions. If these mergers occur within triple or higher-order multiple systems, the resulting BSS can have a companion. But the companion is typically a normal MS star, and the orbit is wide with high eccentricity, which reflects the original orbit configuration of the inner binary and the tertiary.

Accurately determining the multiplicity of BSSs poses a considerable challenge, due to the need for high-quality, multi-epoch spectroscopic observations. Despite these challenges, significant progress has been made in studying low-mass field stars and several OCs, particularly M\,67 and NGC\,188 \citep{2007HiA....14..444L,2008AJ....135.2264G}.
Notably, high binary fractions of approximately $79\% \pm 24\%$ (11 out of 14) and $76\%\pm 19\%$ (16 out of 21) among BSSs were found in M\,67 and NGC\,188, respectively \citep{2015AJ....150...97G,2009Natur.462.1032M}. It is worth noting that these rates are significantly higher than those of normal MS stars in the same clusters. Most of these binaries have long orbital periods, ranging from 700 to 3000 days (here the upper limit is set by observational constraints). The companion mass distribution peaks at approximately 0.5 $M_\odot$ \citep{2011Natur.478..356G,2012AJ....144...54G}, suggesting that they are white dwarfs. \cite{2015ApJ...814..163G} further conducted a far-ultraviolet (FUV) survey with the Hubble Space Telescope (HST) of the BSSs in NGC\,188 and found seven of them have WD companions. This indicates that at least 33\% BSSs in this cluster were formed through binary mass transfer in the past 400\,Myr. Similar fraction of BSSs with WD companions are also found in FUV observations of M\,67 and the younger OC NGC\,7789 (1.6\,Gyr) \citep{2021MNRAS.507.2373P,2022MNRAS.511.2274V}. These findings strongly support a binary mass transfer origin for BSSs in these OCs. Moreover, the existence of several long-period binaries with nearly zero eccentricity and a few double-lined spectroscopic binaries (SB2) with periods around 5 days further reinforce the binary mass transfer scenario. 

For field BSSs, studies such as \cite{2000AJ....120.1014P} have identified a high fraction of binaries, almost four times greater than that of normal field MS stars. Similarly to BSSs in OCs, these binaries generally have orbital periods spanning hundreds of days, with companion masses peaking at approximately $0.6\,M_\odot$. The predominance of single-lined spectroscopic binaries (SB1), in which companions remain undetected, further points to white dwarf companions and a likely binary mass transfer origin. 

Only recently has the binary fraction of BSSs in GCs been measured using spectroscopic data. \cite{10.1093/mnras/stae2333} reported a binary fraction of $13.6\pm5.1$\% for the BSSs in $\omega$ Centauri, which is significantly higher than the $1.2\pm0.1$\% binary fraction for normal MS stars in the same cluster. Similarly, \cite{2024arXiv241213189M} found a binary fraction of $10.9\pm4.8$\% for BSSs in GC 47\,Tucanae, compared to $2.4\pm1.0$\% for normal MS stars. 
Although the measured binary fraction of BSSs in GCs are lower than those observed in OCs, the consistently higher binary fraction among BSSs compared to normal MS stars in both environments strongly suggests that binary evolution plays an important role in BSS formation across different cluster environments. 
In addition, surveys observing variability in GCs have shown that W Ursae Majoris variable (W UMa-type binaries) are common among BSSs \citep{2000AJ....120..319R}. These are low-mass contact binaries that are undergoing mass transfer. The prevalence of W UMa-type binaries among BSSs in GCs suggests that binary mass transfer plays a significant role in their formation even within these dense stellar environments.

It is worth mentioning that the high binary frequency of BSSs observed in some environments may rule out single star evolution through unusual internal mixing as the predominant formation scenario for BSSs (see Section.\,\ref{sec:form}). Additionally, mergers that include only two stars and produce a single star may also be considered less favorable.

\subsection{Rotation}
Rotation significantly impacts stellar structure and evolution through two opposite effects. Centrifugal forces in a rapidly rotating star can deform the star, reducing its effective gravity. According to von Zeipel's theorem ($T_\mathrm{eff}\sim g^{1/4}_\mathrm{eff}$, where $T_\mathrm{eff}$ and $g_\mathrm{eff}$ represent the effective temperature and effective gravity, respectively), this reduction in effective gravity causes a star to appear cooler and less luminous \citep{1924MNRAS..84..665V}. Conversely, rotationally-induced internal mixing can transport fresh hydrogen into the stellar core and bring hydrogen burning products, such as helium, to the surface. This mixing process will make a rapidly-rotating star hotter, more luminous, and having a longer MS lifetime. The properties of rotating stars are determined by the interplay of these two effects \citep{2000ARA&A..38..143M}.

The rotational velocities of BSSs offer critical information into their formation processes. 
BSSs formed through binary mass transfer are typically expected to exhibit fast rotation due to the accretion of angular momentum. In contrast, BSSs resulting from mergers may have varying rotational velocities depending on their evolutionary status, because while the immediate merger products are believed to rotate at near critical velocities, several mechanisms such as magnetic braking and stellar winds can spin them down (see Section\,\ref{sec:form} for details).

Observations indicate that BSSs display a wide range of rotational velocities, from slow to fast compared to background stellar populations \citep{2010ApJ...719L.121L,2013ApJ...772..148L,2023A&A...680A..32B}. As previously mentioned, in the OC NGC\,188, a high binary fraction of BSSs indicates that binary mass transfer is the dominant formation mechanism. Indeed, \cite{2009Natur.462.1032M} found that these BSSs rotate faster than normal MS stars in the same cluster. Recent analysis of 320 BSSs across eight GCs by \cite{2023NatCo..14.2584F} reveals that while normal stars in these clusters generally exhibit negligible rotational velocities, BSSs often rotate faster, with 28\% displaying velocities above 40 $\mathrm{km\, s^{-1}}$.
Notably, this study highlights the influence of environmental density on BSS rotation. In 
less dense clusters like M\,55, M\,4, $\omega$ Centauri and NGC\,3201, about 38\% (87 out of 228) of BSSs have rotational velocities larger than 40 $\mathrm{km\, s^{-1}}$, compared to just 4\% (4 out of 92) in denser clusters such as M\,30, NGC\,6752, NGC\,6397 and 47\,Tucanae (see Fig.\,3). These findings support the notion that binary mass transfer plays a more important role in the formation of BSSs in lower-density environments, consistent with other observations related to their radial distributions and binary statuses.

Unlike old clusters, where normal MS stars have slow rotation due to magnetic braking, young and intermediate-age OCs (younger than $\sim$1\,Gyr) are suggested to contain substantial moderately fast-rotating normal MS stars. Observations from the HST over the past decade have revealed double MSs (a red MS with about 70-90\% stars and a blue MS with about 10-30\% stars) in young and intermediate-age OCs in the Magellanic Clouds \citep{2016MNRAS.458.4368M,2018MNRAS.477.2640M,2017ApJ...844..119L}. The red MS stars are believed to rotate fast, at about half their critical velocities, while the blue MS stars are likely to rotate more slowly, typically below about 30\% of their critical velocities \citep{2009MNRAS.398L..11B,2019ApJ...887..199G,2022NatAs...6..480W}. This bimodal distribution of rotational velocities coincides with findings in field B-type and early A-type stars \citep{2012A&A...537A.120Z,2013A&A...550A.109D}. 
Recently, \cite{2022NatAs...6..480W} suggested that while red MS stars in these clusters are normal MS stars, the blue MS stars might be products of mergers. BSSs in these clusters appear at the brighter end of the blue MS and are also likely products of binary mergers. 

It is still unclear whether merger products have slow or fast rotation. A recent study by \cite{2019Natur.574..211S} suggested that merger products of massive stars exhibit slow rotation once they restore thermal equilibrium (see Sec.\,\ref{sec:form_binary_merger} for details). Spectroscopic studies of individual stars in young OCs NGG\,330 ($\sim$ 40\,Myr old) and NGC\,1846 ($\sim$ 1.5\,Gyr old) show that BSSs have slower rotation (an average value of $v\sin i\sim 50 \kms$) compared to other normal MS stars (an average value of $v\sin i\sim 160 \kms$) \citep{2020MNRAS.492.2177K,2023A&A...680A..32B}. A few BSSs exhibit moderately fast rotations, ranging from 150 to 300 km/s, but these speeds are similar to the fastest rotators among the normal MS stars. These findings support a binary merger origin for BSSs in young OCs. It is unclear how binary mass transfer contributes to BSSs in young OCs.

Although the absolute rotational velocities of BSSs in both young and old OCs might appear similar, their velocities relative to other normal MS stars differ significantly. A comprehensive theoretical study of stellar rotations resulting from various formation mechanisms for both high-mass and low-mass stars is essential to understand the distinct rotational behaviors observed between young and old OCs.

\begin{figure}[t]
\centering
\includegraphics[width=.8\textwidth]{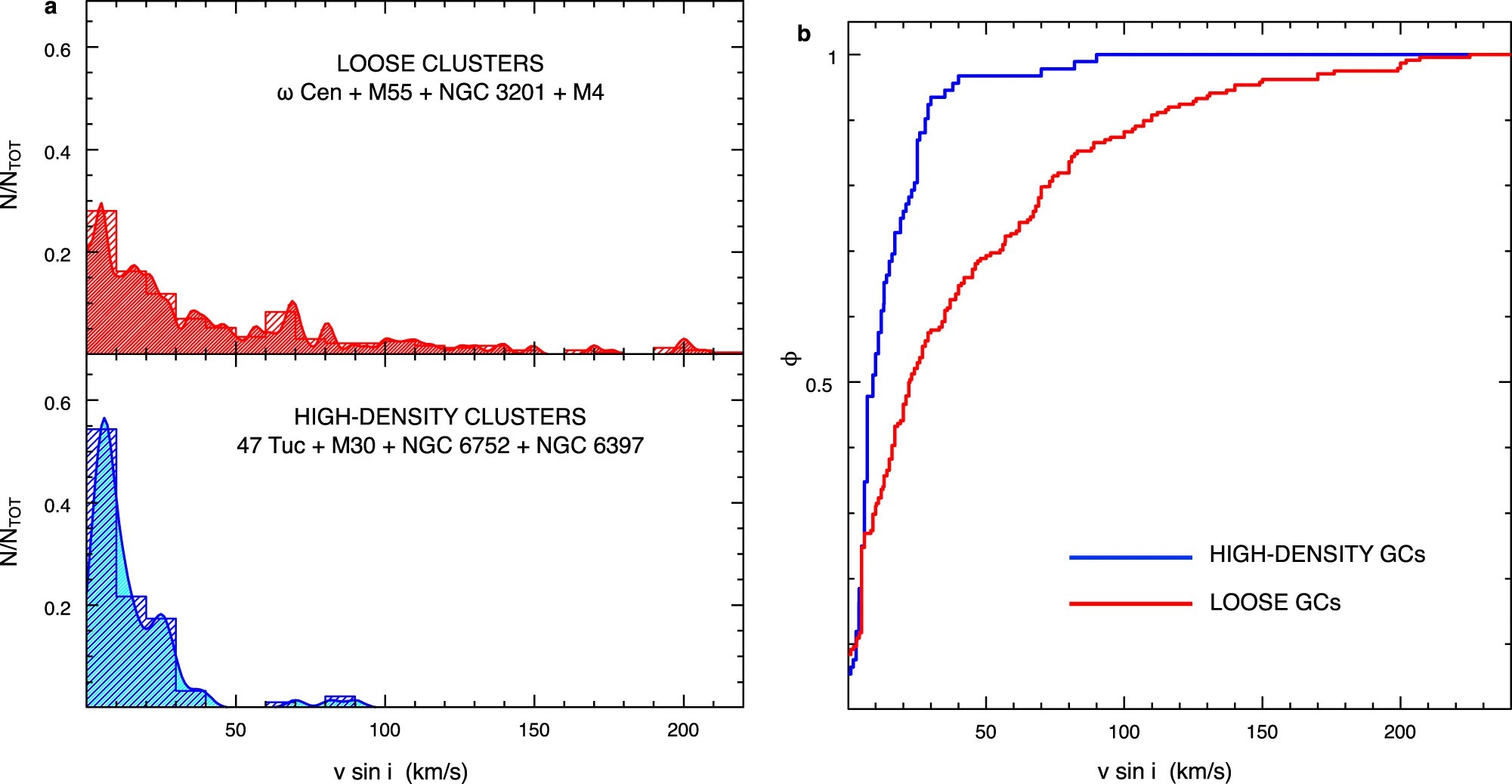}
\caption{Comparing rotational velocities of blue straggler stars in different environments.
a: rotational velocity distribution for blue straggler stars in four loose clusters (upper panel) and four high-density clusters clusters (lower panel). The fraction is obtained by dividing the number of blue straggler stars in each bin by the total number of blue straggler stars in the sample. b: Cumulative distribution of rotational velocities for blue straggler stars in loose clusters (red line) and high-density clusters (blue line). This figure is from Fig. 2 in \cite{2023NatCo..14.2584F}.}
\label{fig:rotation}
\end{figure}

\subsection{Chemical composition}\label{sec:ob_chem}
The surface chemical compositions of BSSs are directly related to their formation processes and evolutionary history. 3D hydrodynamic simulations suggest that BSSs from mergers, including both binary mergers and stellar collisions, typically have surface chemical compositions similar to those of normal MS stars (see Sec.\,\ref{sec:form}). 
In contrast, BSSs arising from binary mass transfer are likely to show abnormal chemical compositions at their surfaces, due to accretion from the deep layers of their evolved donor companions, where the CNO burning occurred \citep{1996QJRAS..37...11S}.

Despite the potential of surface chemical compositions to reveal the origins of BSSs, accurately measuring chemical data remains challenging. A significant discovery in this field came from the GC 47 Tucanae, where a population of BSSs with depleted carbon and oxygen was identified, making the first direct chemical evidence pointing to a binary mass transfer origin in GCs \citep{2006ApJ...647L..53F}. In this study, of the 42 BSSs analyzed, 36 show chemical abundances similar to those of the cluster’s turn-off stars, while 6 display significant depletions in carbon and oxygen. Two of the carbon and oxygen depleted BSSs are in W UMa-type binaries. Notably, these carbon and oxygen depleted BSSs are also among the least luminous BSSs in this cluster, supporting the idea that they are products of binary mass transfer rather than merger events, as merger products are generally more massive and luminous. 
However, it is puzzling that these carbon and oxygen depleted BSSs show similar rotational velocities as other BSSs in the cluster, with only two moderately fast rotators ($v\,\sin i \sim 21$ and $13\,\mathrm{km\, s^{-1}}$). This suggests that the formation mechanisms of BSSs are complex and remain not fully understood. 

Similar complexity has been observed in another cluster GC M\,4. In this cluster, all 14 measured BSSs have normal C and O abundances \citep{2010ApJ...719L.121L}, suggesting a possible merger origin. However, 40\% of them show fast rotation between 50 and 150$\kms$. This fast rotation typically indicates a binary mass transfer origin and challenges the merger hypothesis.

If a BSS is formed through accretion from an asymptotic giant branch (AGB) star, it may exhibit surface enrichment of s-process elements like barium (Ba), which are produced during the thermally-pulsing phase of AGB evolution. This has been observed in BSSs within several OCs, including NGC\,6819 \citep{2015AJ....150...84M}, NGC\,7789 and M\,67 \citep{2024ApJ...970..187N}. However, the absence of Ba enrichment does not necessarily rule out an AGB mass transfer origin, as the mass transfer could occur before the donor star enters the thermally-pulsing phase. Furthermore, the extent of Ba enrichment appears linked to the mass of the AGB donor star, with more massive AGB stars contributing more Ba during dredge-up events \citep{1983ARA&A..21..271I}. This can explain why the fraction of Ba-enriched BSSs decreases with increasing cluster age \citep{2015AJ....150...84M}.


\section{Formation mechanisms}\label{sec:form}
To generate a BSS, mechanisms that add mass to a normal MS star are essential. These mechanisms include binary mass transfer and mergers resulting from binary evolution or stellar collision. Consequently, the formation of a BSS typically requires the involvement of two or more stars. In rare instances, single stars undergoing chemically homogeneous evolution due to rapid rotation and efficient rotationally-induced mixing can remain hot and undergo significant rejuvenation during their MS phase. However, this process predominantly occurs in very massive stars (more massive than approximately $ 20\,M_\odot$) with rotational velocities higher than about 400$\kms$ \citep{2011A&A...530A.115B}. Given its rarity, chemically homogeneous evolution does not account for the majority of observed BSSs and is not considered a primary formation mechanism. This section explores the theoretical aspects of BSS formation, focusing on mechanisms that involve two or more stars. The predicted properties from different mechanisms are summarized in Table\,\ref{tab:2}.

\begin{table}
\begin{center}
\caption{Summary of BSSs models and their prediction. Triple star evolution is considered in binary mergers and stellar collisions, as its main impact is inducing the mergers of inner binaries. }
\label{tab:2}
\begin{tabular}{ l c c c c c c c} 
\toprule
\midrule
Theoretical model property & Binary merger  & Stellar collision & Binary mass transfer \\
\midrule
Binarity  &  Low or high (in triples) & Low or high (in triples)& High\\
Companion  &  No companion  & No companion  & main-sequence stars (Algols), \\
   & or any type companion (in triples) & or any type companion (in triples)& helium stars, compact objects\\
Period & Mostly longer than hundreds of days & Mostly longer than hundreds of days & Short to long \\
Spatial distribution  &  --   & Centrally concentrated  & -- \\
Eccentricity & High & High & Low\\
Surface chemical composition & Normal ? & Normal ? & CNO  \& helium anomalies \\
Rotation & Slow to fast ? & Slow to fast ? & Fast\\
\bottomrule
\end{tabular}
\end{center}
\end{table}

\subsection{Binary mergers}\label{sec:form_binary_merger}

\begin{figure}[t]
\centering
\includegraphics[width=.8\textwidth]{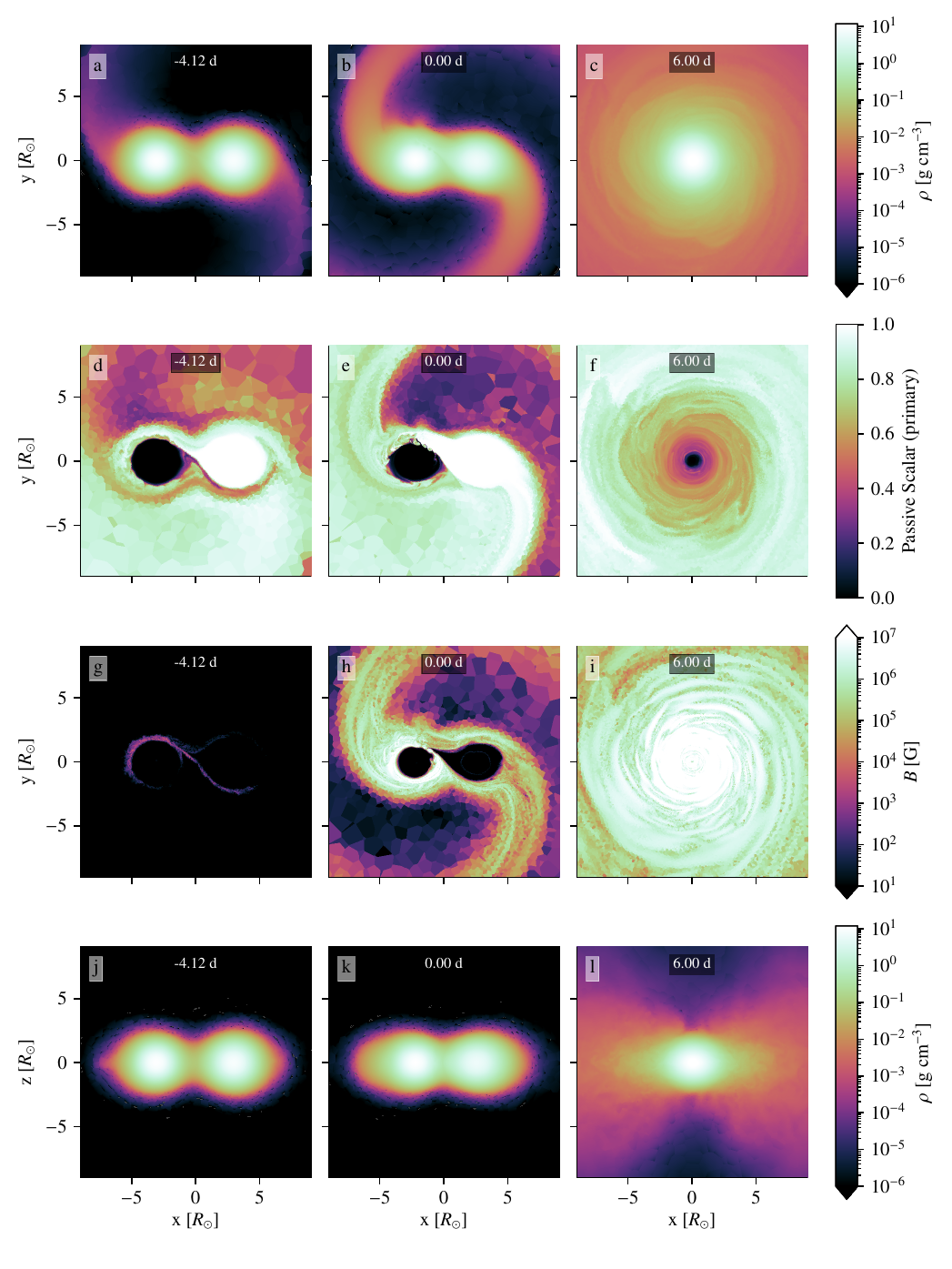}
\caption{Changes in Density, passive scalar and magnetic-field intensity during the 3D magnetohydrodynamic simulation of a merger between a $9\,M_\odot$ and an $8\,M_\odot$ star. Panels a-c: density variations in the orbital plane. Panels d-f: how material from both progenitors mixes during the merger process, with the passive scalar indicating material from the $9\,M_\odot$ primary star. Panels g-i: magnetic-field intensity in the orbital plane. Panels j-l: an edge-on view of the density variation. The time shown in each panel indicates the time elapsed since the moment when the cores of the two stars merged. The figure is reproduced from Fig.\,1 in \cite{2019Natur.574..211S}.}
\label{fig:3d_sim}
\end{figure}

Binary mergers refer to two close binary stars that undergo contact or unstable mass transfer and eventually merge into a more massive star. Both stars must be MS stars because if one star completes central hydrogen burning and forms a helium core, the merger product will inherit this core and become a post-MS star, which does not qualify as a BSS. The mechanisms driving binary mergers differ between low-mass and high-mass stars. In low-mass stars, magnetic braking plays a crucial role in contracting the binary orbit. This is commonly seen in contact binaries, like W UMa-type binaries, which are believed to be merger progenitors. Conversely, high-mass binaries, typically lacking significant magnetic fields due to their radiative envelopes, experience mergers triggered by the expansion of the stars during their evolution.

\cite{2001ApJ...552..664N} conducted a comprehensive study on Case A mass transfer, which occurs when the donor star is still in its MS phase (see Sec.\,\ref{sec:form_MT}), across a wide range of binary systems with primary masses from 0.8 to 50$\mso$. They found that contact between the two stars, potentially leading to mergers, is a common outcome. 
Furthermore, \cite{2024A&A...682A.169H} explored the formation mechanisms and binary parameter space for mergers of two massive stars (with primary masses between 5 and 20$\mso$). They identified several mechanisms that can initiate a merger, such as accretor expansion, $\mathrm{L}_2$ overflow, tidal-induced orbital decay, and run-away mass transfer. Their study shows that in this mass range, approximately 8\% of the binaries will merge when both stars are MS stars and lead to BSSs. However, it is important to acknowledge that these theoretical predictions are subject to many uncertainties regarding massive binary evolution, such as the initial binary parameter distribution \citep[][and references therein]{2017ApJS..230...15M} and mass transfer stability \citep{2015ASSL..413..179I,2024ARA&A..62...21M}.

The merger of two stars is inherently a 3D process that cannot be adequately modeled by 1D stellar evolution code. In 1D simulations, simple assumptions, such as complete rejuvenation or complete mixing need to be made. With advancements in computational power, 3D simulations that can illustrate the internal structure, rotation profile, mass loss during the merger event and magnetic fields of merger remnants have been feasible. \cite{2019Natur.574..211S} conducted the first and only 3D magnetohydrodynamic (MHD) simulations to date of binary mergers involving two massive MS stars using the moving-mesh code AREPO \citep{2010MNRAS.401..791S}. They studied the merger of a binary system consisting of a 9$M_{\odot}$ and a 8$M_{\odot}$ MS star that undergoes dynamically unstable mass transfer (see Fig.\,\ref{fig:3d_sim}). Their aim was to explain the observed properties of $\tau$ Sco \citep{2009ARA&A..47..333D,2014A&A...566A...7N,2016MNRAS.457.2355S}, a massive star ($\sim$15$\mso$) with strong surface magnetic fields that appears young (the inferred age is $<5$\,Myr) compared to other presumably coeval members of the Upper Scorpius association ($\sim 11$\,Myr). 

Their simulation showed that the core of the merger product primarily consists of material from the secondary star, while the torus is composed mainly of material from the primary star. The most important result of this 3D simulation is that strong surface magnetic fields (thousands of G) can be generated by the magneto-rotational instability in the merger remnant. The magnetic fields may be long-lived, as the Ohmic decay timescale is comparable to stellar lifetime. This discovery provides new insights into the origins of magnetic fields in massive stars. Unlike Sun-like stars, which generate magnetic fields through a dynamo process within their convective envelopes, intermediate-mass and high-mass stars primarily have radiative envelopes where such a dynamo process is typically absent. Despite this, observations indicate that approximately 10\% of massive stars possess strong magnetic fields \citep{2013MNRAS.429..398P}, the origins of which have been unclear. 

\begin{figure}[t]
\centering
\includegraphics[width=.8\textwidth]{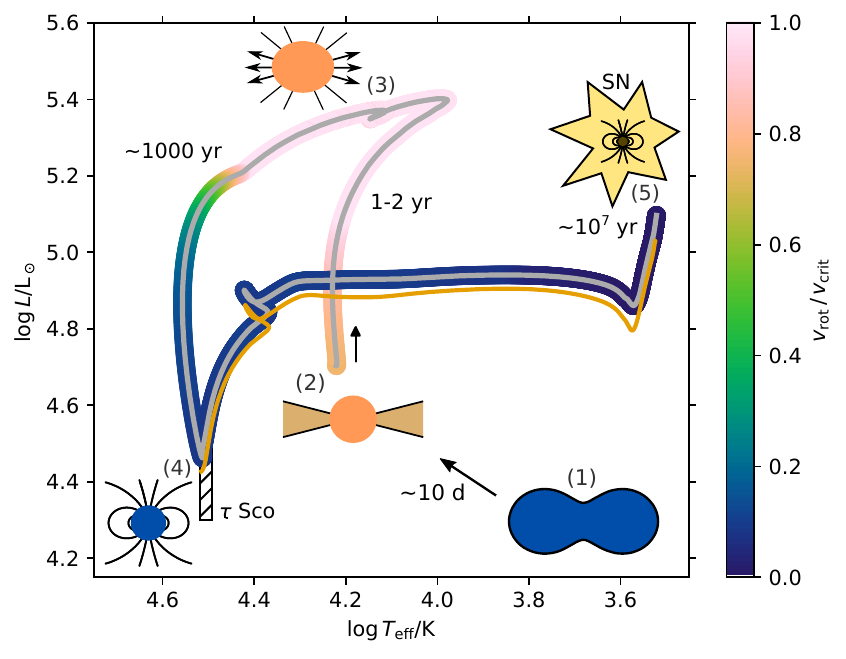}
\caption{Comparison of the long-term evolution of the merger product and a genuine single star in the Hertzsprung-Russell diagram. The grey line shows the evolution of the merger product, with shading color-coded by the surface rotational velocity in terms of the critical Keplerian velocity. The orange line depicts the evolution of a genuine single star with $16.9\,M_\odot$. The small cartoons correspond to key evolutionary phases of the merger remnant: (1) contact phase before coalescence, (2) torus-core structure of the merger remnant, (3) inflated star that achieves critical rotation, (4) the star settling back into thermal equilibrium with strong magnetic fields, (5) the star eventually explodes as a supernova. The figure is reproduced from Fig.\,2 in \cite{2019Natur.574..211S}.}
\label{fig:merger_ev}
\end{figure}

3D simulations are well-suited for modeling dynamic processes but are not suitable for long-term stellar evolution studies. For this purpose, \cite{2019Natur.574..211S} employed the 1D stellar evolution code MESA \citep{2011ApJS..192....3P,2013ApJS..208....4P,2015ApJS..220...15P} to simulate the subsequent evolution of the merger remnant. The outcome of the 3D simulation indicated that the remnant is inflated and initially rotates nearly at the critical velocity. 
They found that as the star contracts to regain thermal equilibrium over the subsequent several thousand years, its surface spins down significantly, to about 10\% of critical rotational velocity. This is due to the internal restructuring of the star, with the envelope contracting and the core expanding. 
Magnetic braking may further spin down the star. Eventually, when the star stabilizes back to thermal equilibrium, it exhibits slow rotation and has almost the same properties as a genuine single star of the same mass (see Fig.\,\ref{fig:merger_ev} for the evolution of the binary merger product).

The case study by \cite{2019Natur.574..211S} underscores the feasibility and crucial importance of such 3D simulations in BSS research. However, such studies are still in their early stages, primarily because 3D simulations require substantial computational resources.
Future studies that incorporate a wider range of initial binary parameters, particularly including low-mass stars, are essential for a more comprehensive understanding of BSSs across various environments.

\subsection{Stellar collisions}\label{sec:form_collision}
Stellar collisions are violent encounters between two or more stars, which can result in mergers. These events are frequent in densely populated environments, such as GCs and the cores of OCs. Since the study of \cite{1976ApL....17...87H}, it has been widely accepted that stellar collisions play an important role in forming BSSs in GCs. While stellar collisions also contribute to the formation of BSSs in OCs \citep{1999A&A...348..117P,2001MNRAS.323..630H}, their impact is generally less pronounced compared to GCs. Stellar collisions can occur between two single stars or within higher-order multiples, which are perhaps more frequent due to their larger collisional cross sections \citep{1991AJ....102..994L,2004MNRAS.352....1F,2013ApJ...777..106C}. In extremely dense environments, such as young compact clusters, repeated collisions may happen and lead to the so-called `merger runaway' \citep{1999A&A...348..117P}.

Unlike binary mergers, where the relative velocity of the two progenitor stars is dictated by their binary orbital period, stellar collisions can occur between stars with a much wider range of relative velocities. This difference may result in distinct structures of the merger remnants in these two cases. 

Stellar collisions and their merger products have been studied using 3D smoothed particle hydrodynamic (SPH) simulations \citep[e.g.][]{1987ApJ...323..614B,1996ApJ...468..797L,1997ApJ...487..290S,2013MNRAS.434.3497G,2005MNRAS.358..716S}. These 3D simulations allow us to study the mass loss during stellar collision, the internal structure and rotational velocity profile of the merger remnants \footnote{Performing 3D simulations is time-consuming. To address these challenges, \cite{2002ApJ...568..939L} developed a tool called `Make Me A Star', which employs simplified algorithms to model the structural outcomes of stellar collision based on entropy sorting demonstrated in 3D SPH simulations. This tool has been integrated with N-body calculations in population synthesis studies for BSSs \citep{2008A&A...488.1007G}.}. These simulations demonstrate that the mass loss during stellar collision is determined by the collision parameters and mass ratios of the merger progenitors, and is usually less than about 10\% of the total mass \citep{2013MNRAS.434.3497G}. The structure of the collisional merger products adheres to the entropy sorting principle, where entropy increases from the center to the surface \citep{1996ApJ...468..797L}. Typically, for unevolved stars, the core of the less massive star sinks to the center, while the more massive star forms the envelope. In contrast, for evolved stars,  entropy decreases faster in more massive stars, and as a result, the core of the more massive primary star sinks to the center of the merger remnant.

Mixing plays a crucial role in the evolution of merger remnants, as it introduces fresh hydrogen into the core. Enhanced mixing can extend the MS lifetime of the resulting BSSs and lead to higher luminosity and temperature. 3D simulations reveal that, in general, little hydrogen is mixed into the core for both low-mass and high-mass stars, but significant nitrogen enhancement is seen at the surface of the merger product. Interestingly, this predicted nitrogen enhancement agrees well with the values found in magnetic early B-stars which could be merger products \citep{2008A&A...481..453M}. In off-axis collisions, simulations predict a slightly higher degree of mixing compared to head-on collisions \citep{2001ApJ...548..323S}. 

The rotational velocity of merger products can significantly vary based on the dynamics of the collision. 3D simulations show that while the internal structure of merger products, such as density, temperature, and chemical composition, does not strongly depend on the collision’s impact parameter, the rotational velocity does. Specifically, off-axis collisions, where the colliding stars have initial tangential velocities relative to each other, produce remnants with higher rotational velocities than head-on collisions. This is because the tangential velocities contribute to the system’s total angular momentum, influencing the resulting spin of the merger product.
In off-axis collisions, the merger remnants often retain too much angular momentum and reach critical rotational velocities. This excess angular momentum can prevent them from contracting and settling into the MS phase. Significant angular momentum may need to be shed to allow the star to regain thermal equilibrium, possibly through mechanisms such as magnetic torques between the disk or wind and the merger remnant \citep{2001ApJ...548..323S,2005MNRAS.358..716S}. It remains uncertain whether merger remnants from stellar collisions will undergo spin-down due to internal restructuring, as observed in remnants from binary mergers (see Sec.\,\ref{sec:form_binary_merger}). To address this, comprehensive studies are necessary to investigate the thermal relaxation processes in stars of varying masses and initial rotational velocities, along with the potential influences of magnetic fields.

The studies on the long-term evolution of merger products from stellar collisions are conducted using 1D stellar evolution codes. These studies indicate that their subsequent evolution is greatly influenced by factors such as stellar rotation, mixing, and the transport of internal angular momentum, all of which are still uncertain. When strong rotation and mixing are present, the MS lifetimes and evolutionary tracks of the resultant BSSs can significantly differ from those of normal stars \citep{1997ApJ...487..290S,2001ApJ...548..323S}. Conversely, in scenarios like head-on collisions, where stellar rotation and mixing are minimal, the evolution of the merger product tends to resemble that of a regular single star of equivalent mass \citep{2013MNRAS.434.3497G}.

3D SPH simulations have been performed for both low-mass and high-mass stars \citep{2007ApJ...668..435S,2008A&A...488.1007G,2013MNRAS.434.3497G}. While stellar collisions among low-mass stars are common in old GCs, collisions involving massive stars can also occur, particularly in young compact clusters \citep{2007ApJ...668..435S}. Such events are crucial for the formation of intermediate-mass black holes in star clusters \citep{2024Sci...384.1488F}. However, 3D MHD simulations, which are essential for tracking the generation of magnetic fields and have been employed to study binary mergers, are still limited for stellar collisions. Pioneering work has been conducted by \cite{2024arXiv241000148R}. There is an urgent requirement for such studies to enable a comprehensive comparison of the structural and evolutionary outcomes of mergers arising from stellar collisions and binary mergers.

\subsection{Binary mass transfer}\label{sec:form_MT}
Binary mass transfer occurs when one star in a binary system (known as the donor star) fills its Roche lobe (i.e., the critical equipotential surface through the first Lagrangian point $L_1$) and transfers material to its companion (known as the accretor). This phenomenon was first calculated by \cite{1960ApJ...132..146M} in an effort to explain the so-called \textit{Algol paradox} \citep{1955ApJ...121...71C}.
\cite{1964MNRAS.128..147M} was the first to propose that BSSs could form through this process. Notably, \cite{2021ApJ...908....7S} and \cite{2023ApJ...944...89S} found that the blue straggler binaries WOCS 5379 and WOCS 4540 in NGC\,188 are well explained by models of binary mass transfer. Extensive studies on binary mass transfer are documented in several works \citep{1966AcA....16..231P,1967AcA....17..355P,1967ZA.....65..251K,1991A&A...241..419P,1992ApJ...391..246P,1994A&A...290..129V,1998A&ARv...9...63V,2004MNRAS.355.1182C,2006epbm.book.....E,2012ARA&A..50..107L,2013A&ARv..21...59I,2023pbse.book.....T}.

Binary mass transfer is classified into different types based on when the donor star fills its Roche lobe:
\begin{itemize}
\item \textbf{Case A}: mass transfer occurs when the donor star is burning hydrogen in the core.
\item \textbf{Case B}: mass transfer occurs after the donor star has completed central hydrogen burning but before it begins central helium burning.
\item \textbf{Case C}: mass transfer occurs after central helium depletion.
\end{itemize}

The concepts of Case A and Case B mass transfer were introduced by \cite{1967ZA.....65..251K}, and the Case C mass transfer was later introduced by \cite{1970A&A.....7..150L}. Because stars expand during evolution, this classification means that the initial orbital periods required for Case A, B, and C mass transfers increase sequentially. 

Unlike mergers that occur on dynamical timescales and can be simulated using 3D methods, stable mass transfer happens on nuclear or thermal timescales. Consequently, 1D stellar evolution codes are employed to compute binary mass transfer. When computing binary evolution, mass transfer stability and efficiency are critical yet poorly understood factors \citep{2007A&A...467.1181D,2008MNRAS.384.1109E,2012ARA&A..50..107L}. Stable mass transfer can rejuvenate a star, potentially leading to a BSS, while unstable mass transfer tends to result in a merger, which only forms a BSS if both progenitors are MS stars. The type of mass transfer does not directly predict stability, instead, it is influenced by the donor’s structure, which is crucial for determining the star's response to mass loss, and therefore, mass transfer stability \citep{2015ASSL..413..179I}. Generally, for massive stars, Case A and B mass transfers, where the donor has a radiative envelope, are more likely to be stable. Case C mass transfer occurs when the donor star has developed a deep convective envelope, often leading to rapid expansion as it loses mass. This can result in the star quickly overfilling its Roche lobe, initiating unstable mass transfer. However, in low-mass stars, Case C mass transfer can still proceed stably \citep{2007AIPC..948..431C,2008MNRAS.387.1416C,2015ASSL..413..179I}.

Regarding mass accretion efficiency, it is unclear how much of the transferred material successfully ends up on the accretor \citep{1992A&AS...96..653D,2002MNRAS.335..948C,2004MNRAS.355.1182C,2007A&A...467.1181D,2008MNRAS.384.1109E,2012ARA&A..50..107L,2014ApJ...796...37S}. 
A pivotal question is whether the accretor can continue accreting after reaching critical rotation. Historically, models often neglected the spin-up of accretors and assumed constant mass transfer efficiency. Recent approaches consider accretor rotation, stopping accretion when it reaches critical rotation. These recent studies suggest that shorter-period binaries, where tidal forces can efficiently spin down the accretor, exhibit higher mass transfer efficiency. Whereas long-period Case B binaries have highly non-conservative mass transfer \citep{2022NatAs...6..480W}. 
For BSS formation via mass transfer, the accretor must gain enough mass to exceed the cluster’s turn-off mass \citep{2021ApJ...908....7S,2023ApJ...944...89S}. Under the rotation-restricted mass transfer efficiency assumption, most binaries are unlikely to produce BSSs, as a star can reach critical rotation by accreting only tiny amounts of material \citep{1981A&A...102...17P}. However, some studies propose that accretors might continue to accrete mass even after reaching critical velocities by transporting angular momentum outwards via accretion disks \citep{1991ApJ...370..604P}. This mechanism could facilitate the formation of BSSs. 
It is worth noting that recent studies have identified a population of so-called blue lurkers. These stars may undergo similar formation mechanisms as BSSs, such as mass accretion, but are not massive enough to surpass the cluster turn-off point \citep{2019ApJ...881...47L,2023ApJ...944..145N,2024ApJ...969....8S}.

The most distinct features of BSSs resulting from binary mass transfer are their typically fast rotation and abnormal surface chemical compositions. The exact rotation may be influenced by the initial orbital periods of their progenitor binaries. BSSs originating from Case A mass transfer can have either slow or fast rotation, depending on the efficiency of tidal interactions that spin them down \citep{2020ApJ...888L..12W,2022A&A...659A..98S}. Case B mass transfer typically leads to fast-rotating BSSs. The BSSs formed through Cases A and B mass transfer might exhibit unusual surface carbon and oxygen abundances. Case C mass transfer usually results in long-period binaries (of order 1000 days) and may include s-processed materials such as barium if the donor is an AGB star (see Sec.,\ref{sec:ob_chem}).

Another distinct feature of BSSs resulting from binary mass transfer is the presence of companion stars.
The nature of the companion is influenced by the characteristics of the progenitor binary \citep{2019ApJ...878...49W}. For instance, the primary stars more massive than about 30$\mso$ typically evolve into black holes. The primary stars between about 13 to 30$\mso$ typically lead to neutron stars, in which case their binaries may be disrupted due to supernova kicks, and leave a runaway single BSS. The primary stars less massive than 13$\mso$ usually evolves into white dwarfs. Consequently, in younger OCs (younger than about 30 Myr), BSSs from binary mass transfer are likely to have black hole companions, whereas in older clusters, white dwarf companions are more common. The existence of these white dwarf companions among BSSs in OCs and field stars is a strong indicator of their binary mass transfer origin (see Sec.,\ref{sec:ob_multiplicity}).


\subsection{Triple evolution}

The existence of a third star can profoundly influence the evolution of inner binaries \citep{2009ApJ...697.1048P,2016ComAC...3....6T}, often increasing the likelihood of mergers that form BSSs. Triple star evolution was largely ignored in previous studies but has gained more and more attention recently, because observations reveal a high prevalence of triple systems, especially among massive stars \citep{2006A&A...450..681T}. 

Triple evolution can lead to the merger of the inner binary by the so-called Kozai-Lidov mechanism \citep{1962AJ.....67..591K,1962P&SS....9..719L}, in which the distant tertiary induces long-term periodic oscillations in the eccentricity and inclination of the inner binary's orbit, and subsequent tidal friction may ultimately lead to a merger between the inner binary stars. In particular, \cite{2014ApJ...793..137N} demonstrated that the revised eccentric Kozai-Lidov mechanism, where the canonical assumptions of a circular outer orbit and a point-particle inner binary component are relaxed, plays a significant role in the formation of BSSs. Predictions from this revised mechanism aligns well with observations of BSSs in NGC\,188 \citep{2012AJ....144...54G}.
Recently, \cite{2024arXiv240706257S} conducted a detailed exploration of mergers within inner binaries of triple systems, suggesting that triple dynamics likely contribute to 20\% to 25\% of the BSS population. In particular, the widest ($P>1000$ days) and most eccentric BSSs are most likely to result from triple-star evolution. In addition, it has been hypothesized that the third companion could actively transfer mass to the inner binary, potentially triggering a common envelope evolution that leads to a merger and thereby producing a BSS \citep{2009ApJ...697.1048P}. 
BSSs formed through the merger of the inner binary in a triple system should typically have a companion, mainly a MS star, in a long distance, although future dynamical evolution may destroy some very long period binaries. Furthermore, \cite{2022MNRAS.515L..50V} found that Kozai-Lidov oscillations can induce mergers in quadruple systems, potentially resulting in a triple system containing a blue MS star. This mechanism may explain the 
massive triple TIC 470710327 detected by NASA's \textit{Transiting Exoplanet Survey Satellite} (TESS).


Triple star systems appear to play a significant role in the formation of BSSs and offer a potential explanation for the discrepancies between the predicted and observed numbers of BSSs in OCs (see Sec.\,\ref{sec:simu}). However, our understanding of triple star evolution remains in its early stages but is continuously evolving. To better understand the role of triple star evolution in the formation of BSSs, further detailed investigations are essential.

\subsection{Distinguish different formation mechanisms}
While the aforementioned mechanisms are often important in specific environments, they can coexist within a single environment. Distinguishing these mechanisms is crucial, yet challenging, because measuring detailed properties of BSSs is difficult, and because different formation processes can lead to BSSs with overlapping characteristics.

One of the most compelling pieces of evidence supporting a binary mass transfer origin is the presence of companion stars. Many BSSs are found in binaries with periods ranging from hundreds to a thousand days, and companion mass distributions peaking at about $0.55\,M_\odot$, indicating an unseen white dwarf companion (see Sec.\,\ref{sec:ob_multiplicity}). Ultraviolet observations have confirmed the presence of white dwarf companions among BSSs, but more such observations are needed to fully understand their prevalence. In triple systems, mergers can also lead to the formation of BSSs with binary companions. Typically, these companions are MS stars, and the binaries may exhibit high orbital eccentricities.

Fast rotation is also a strong indicator of a binary mass transfer origin for BSSs. However, as discussed earlier, binary mass transfer can also produce slowly rotating BSSs if tidal synchronization is effective. Additionally, off-axis collision mergers might exhibit fast rotation. The mechanisms responsible for spinning down these merger products remain poorly understood.

Unusual surface compositions, such as carbon/oxygen depletion in BSSs, are good indicators of a binary mass transfer history. However, depending on when mass transfer occurs, significant abnormal surface compositions may or may not be observed in BSSs, given current observational limits. 

Additionally, BSSs from binary mass transfer are expected to occupy the low luminosity part of the CMD, close to the cluster turn-off, since mass transfer is not as efficient at adding mass as merger events. 

Combining different observed properties is an effective way to distinguish between the formation mechanisms of BSSs. For example, in GC\,47 Tucanae, carbon and oxygen depleted BSSs, which also fall into a lower luminosity group and include two found in W UMa-type binaries, support their formation via binary mass transfer. However, the fact that these carbon and oxygen depleted BSSs exhibit rotational velocities similar to other BSSs challenges the binary mass transfer hypothesis.

\begin{figure}[t]
\centering
\includegraphics[width=.7\textwidth]{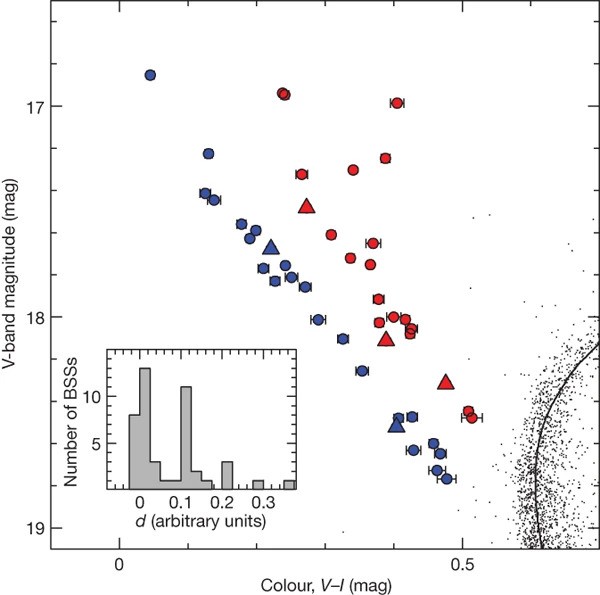}
\caption{The distribution of the two blue straggler sequences in the globular cluster M\,30 in the color-magnitude diagram. Blue and red markers represent the blue and red blue straggler sequences, respectively. Triangles are the detected variables. The inset illustrates the distribution of the distance $d$ of blue straggler stars from the best-fitting straight line to the blue sequence. The figure is reproduced from Fig.\,4 in \cite{2009Natur.462.1028F}.}
\label{fig:double_BSS}
\end{figure}

The identification of double sequences of BSSs in GC M\,30 provides a unique opportunity to differentiate between the mechanisms of binary mass transfer and stellar collision, as shown in Fig.\,\ref{fig:double_BSS} \citep{2009Natur.462.1028F,2019A&A...621L..10P}. These studies suggest that the double sequences are indicative of multiple origins: the blue BSSs likely formed from stellar collisions during a violent core collapse around 1-2 Gyr ago, while the red BSSs are believed to be products of binary mass transfer.
During cluster core collapse, there is a significant enhancement in the formation of collisionally induced BSSs, and their positions in the CMD should follow one sequence due to similar origin times. In contrast, binary mass transfer occurs throughout the entire cluster evolution, resulting in a more spread distribution. 
However, it is important to note that there are two W UMa contact binaries among the blue BSSs, which contradicts the collision origin hypothesis. In addition, \cite{2018ApJ...856...25L} identified similar double sequences of BSSs in the young GC NGC\,2173 (approximately 1.6\,Gyr), located in the Large Magellanic Cloud. Notably, this cluster has a much lower central density compared to M\,30. Furthermore, around half of BSSs in the blue sequence are located in the outskirts of this cluster. These observations challenge the hypothesis that the blue sequence originate from collisions triggered by cluster core collapse. In summary, the origin of the double sequences of BSSs remains elusive. Further detailed investigations are essential to better understand the properties of BSSs and their diverse formation scenarios.

\section{Simulations and comparison with observations}\label{sec:simu}

As discussed above, the formation and evolution of individual BSSs have been analyzed using 3D and 1D codes. However, to compare theoretical predictions with observations, comprehensive population synthesis studies are necessary. These studies simulate the properties of a population of stars, incorporating empirical distributions of initial parameters.

\begin{figure}[t]
\centering
\includegraphics[width=\textwidth]{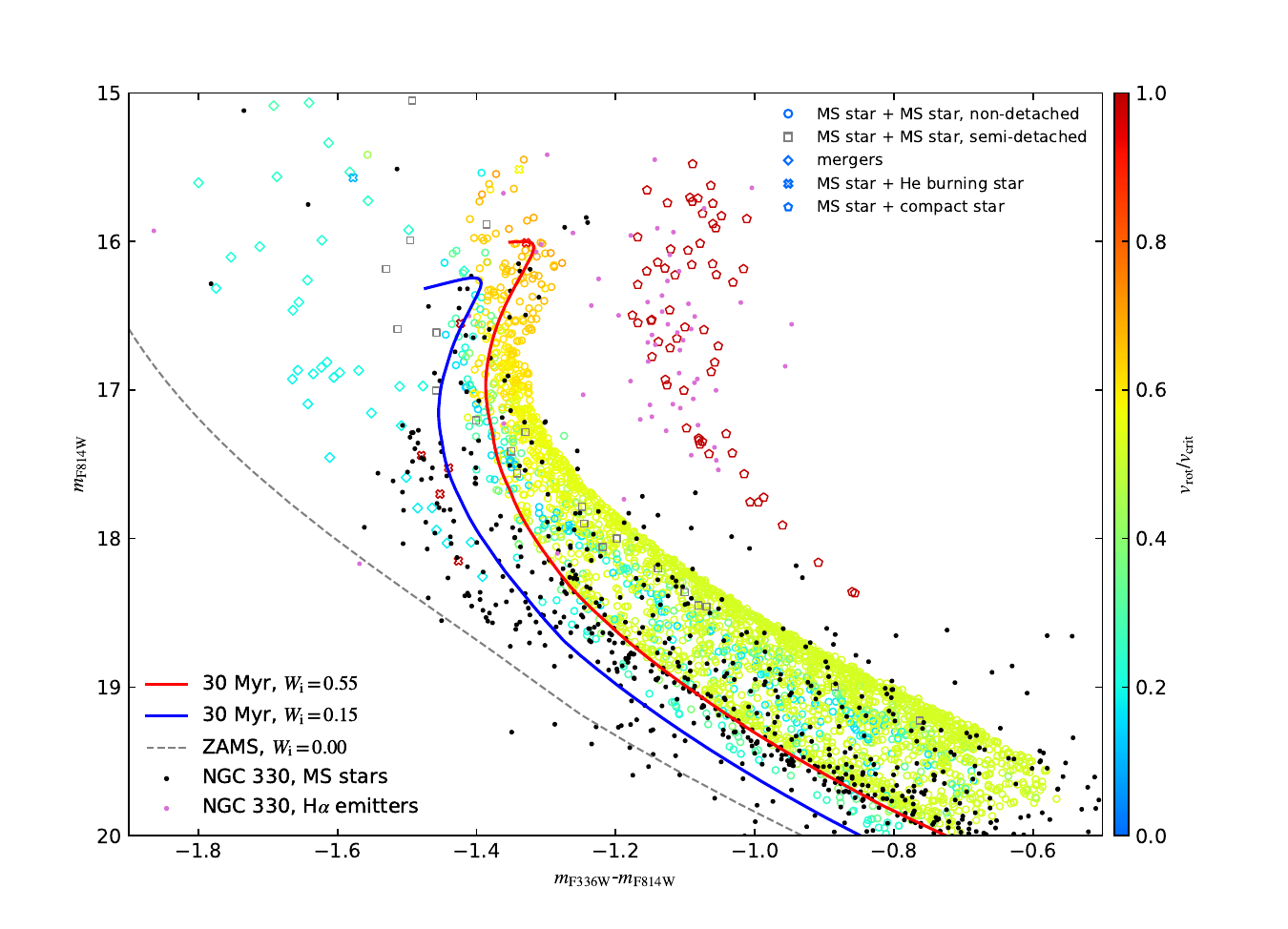}
\caption{Comparison between binary evolution models at 30\,Myr and observed main-sequence stars in NGC\,330 in the color-magnitude diagram. Open symbols represent different binary models or merger products: detached binaries with two main-sequence stars (circles), semi-detached binaries (squares), main-sequence merger products (diamonds), binaries with a main-sequence and a stripped helium star (crosses), and binaries with a main-sequence star and a compact object (pentagons). Semi-detached systems are shown in grey, and the symbol colors indicate the current rotational velocity of the brighter binary component or the merger product. Observed main-sequence stars and H$\alpha$ emitters are overlaid as small black and purple dots, respectively. The red and blue solid lines are isochrones for single star models rotating at 55\% and 15\% of critical rotation initially. The grey dashed line shows the zero-age main sequence for non-rotating stars. It can be seen that main-sequence merger products are likely the primary origin of blue straggler stars in young open clusters. The figure is adapted from Figure\,3 in \cite{2022NatAs...6..480W}. The figure shows a snapshot at 30\,Myr. An animation depicting such simulation up to 100\,Myr \citep{2024ApJ...975L..20W} is available in the online resources.}
\label{fig:NGC330_model}
\end{figure}

Population synthesis simulations using detailed binary models have been carried out to study MS stars in young OCs \citep{1998A&A...334...21V,2020ApJ...888L..12W,2022NatAs...6..480W}. These studies suggest that in young open clusters, BSSs primarily originate from binary mergers (see Fig.\,\ref{fig:NGC330_model}). Moreover, the predicted distribution of BSSs in the CMD aligns well with observations of BSSs in the Small Magellanic Cloud OC NGC\,330.

However, population synthesis using detailed models is computationally expensive, so most current simulations rely on rapid stellar evolution models, such as SeBa \citep{1996A&A...309..179P}, BSE \citep{2002MNRAS.329..897H} and StarTrack \citep{2008ApJS..174..223B,2005MNRAS.358..572I}. Rapid stellar evolution uses a parameterized approach based on pre-calculated stellar models. These population synthesis simulations are complemented by N-body simulations that account for the dynamical evolution of stars within star clusters. Commonly used N-body codes include NBODY6/NBODY7 \citep{2012MNRAS.422..841A,2015MNRAS.450.4070W} and MOCCA \citep{2013MNRAS.429.1221H}. 

The integration of rapid stellar evolution codes with N-body simulations enables researchers to explore the collective properties of BSSs, such as their frequency from different formation channels and their multiplicity, and compare with observations \citep{2013ApJ...777..106C,2001MNRAS.323..630H,2005MNRAS.363..293H}. However, it is still far to compare results from these simulations with individual observed BSSs, because rapid codes rely on simplified assumptions about binary evolution. For example, BSE code assumes complete mixing of merger products, which has been challenged by detailed 3D simulations (see Sections \,\ref{sec:form_binary_merger} and \ref{sec:form_collision}). Further enhancements to these rapid stellar evolution codes, informed by insights from detailed 3D and 1D studies, are in high demand.

Simulations for BSSs in GCs generally align well with observations \citep{2013ApJ...777..106C,2013MNRAS.429.1221H}. These simulations reveal that in dense GCs, most of the BSSs are products of stellar collisions (mostly from binary-single and binary-binary collisions). The number of BSSs predicted by these simulations typically matches the observed counts in GCs. In addition, these simulations report a large fraction of BSSs in binaries with wide orbits, ranging from a few hundred to a few thousand days, which also match observations. However, the observed number of short-period eclipsing BSSs is much higher than that predicted by simulations. This suggests that the role of binary mass transfer in the formation of BSSs in GCs needs further investigation. 

In contrast, simulations for BSSs in OCs show significant discrepancies with observations. \cite{2005MNRAS.363..293H} and \cite{2012AJ....144...54G} have conducted simulations for BSSs in the well-studied clusters M\,67 and NGC\,188. 
These simulations found that the predicted number of BSSs is less than a third of those observed, and the spatial distribution is more centrally concentrated than observed. 
Moreover, while simulations suggest that only about 15\% of BSSs should have binary companions, observational data indicate a much higher frequency of about $\sim 70\%$. 
Although simulations can reproduce the long periods and companion mass distributions of BSS binaries, they primarily result from Case C mass transfer. This process may not transfer enough mass efficiently to form BSSs exceeding the cluster turn-off mass. Therefore, significant tension between theoretical models and observations still exist.

\section{Post-main-sequence evolution of BSSs}\label{sec:later}
In this section, we briefly discuss whether the post-MS evolution of BSSs exhibits distinct properties compared to normal stars. BSSs resulting from mergers have analogous structure and evolution as genuine single star models with equivalent mass, both in the MS phase and the post-MS phase. However, the evolutionary descendent of BSSs can produce distinct features in the CMD. The reason is that due to the rejuvenation process, BSSs commence their post-MS evolution later than their single star counterparts. For instance, when massive BSSs in young OCs evolve to helium-burning red supergiants, they will manifest as red stragglers, i.e. red supergiants that are more luminous than normal red supergiants in the same cluster due to their higher masses \citep{2016MNRAS.458.3968P,2020A&A...635A..29P,2019A&A...624A.128B}. In the case of low-mass stars, the brightest and reddest horizontal branch stars might be progenies of BSSs \citep{1988ARA&A..26..199R,1992AJ....104.1831F,2009ApJ...692.1411S}. Unfortunately, although red stragglers and red horizontal branch stars have been observed, their connection to merger remnants remains unclear.

In contrast, BSSs arising from binary mass transfer may have different evolutionary trajectories compared to genuine single stars. During accretion, whether the accretor's core can expand accordingly to match that of a comparable single star is sensitive to stellar internal mixing, which is still poorly understood \citep{1995A&A...297..483B}. For example, for massive stars, inefficient internal mixing can prevent the accretor's core from expanding adequately, leading to a smaller core-to-envelope ratio compared to a genuine single star. Under such conditions, the accretor may initiate helium burning in the blue supergiant phase \citep{1995A&A...297..483B,2014ApJ...796..121J}. In contrast, a normal single star of similar initial mass would typically evolve into a red supergiant directly.

\section{Outlook}\label{sec:sum}

The sections above provide an overview of our current understanding of BSSs, including their observed properties and formation mechanisms. Research on BSSs spans several disciplines, and significant progress has been made, particularly with the use of 3D simulations in studying binary mergers. However, many questions remain unanswered, necessitating further investigation. This is particularly true for bridging the gap between high-mass BSSs in young OCs and low-mass BSSs in old clusters, especially as recent spectroscopic observations of massive BSSs in young OCs have shown significant progress. In the following, we highlight several promising research directions for the near future.

\begin{itemize}
\item The first 3D MHD simulation of the evolutionary merger of two massive stars is crucial for our understanding of merger products, revealing that strong magnetic fields can be generated. Extending this research to larger binary parameter ranges, especially in the low-mass regime, is an important direction for the near future. 
\item Currently, 3D SPH simulations are the primary tool for studying stellar collisions. In the future, conducting 3D MHD simulations for stellar collisions is vital. This should be done for both low-mass and high-mass stars, enabling a systematic study of the properties of BSSs from both stellar collisions and binary mergers across different mass ranges.
\item Considering both stellar mergers and collisions, one of the most important questions that remains unresolved is the evolution of stellar rotation. The study by \cite{2019Natur.574..211S} shows that while immediate merger products have near-critical rotation, they spin down significantly during the thermal relaxation phase. It is unclear whether this spin-down during thermal relaxation is sensitive to mass and whether it applies similarly to low-mass stars. For low-mass stars, it is also unclear whether magnetic braking plays a more significant role in spinning down merger products. Future systematic studies on the thermal relaxation phase for both high- and low-mass merger products, from both stellar collisions and binary evolution, are needed.
\item Compared to mergers, more uncertainties exist for binary mass transfer, especially regarding its stability and efficiency. As a result, despite understanding how BSSs could form through binary mass transfer, theoretical studies are still not capable of precisely predicting the numbers and detailed properties of BSSs through this scenario across different environments. Enhancing our understanding of binary mass transfer requires studies from various aspects of different binary evolutionary products, not just BSSs.
\item Current population synthesis studies fail to explain the observed number of BSSs in OCs, suggesting that improvements are needed. Including triple star evolution, which is largely overlooked in existing studies, might help address this discrepancy.
\item There is a strong demand for detailed spectroscopic observations that provide a comprehensive characterization of BSSs in exemplary clusters, including multiplicity, rotation, chemical composition, and especially magnetic fields. For instance, in the young OC NGC\,330, while binarity and rotation of BSSs have been well measured \citep{2021A&A...652A..70B,2023A&A...680A..32B}, their chemical compositions and magnetic fields remain unmeasured. Similarly, in the GC M\,30, although an interesting feature of double sequences of BSSs is observed, the detailed properties of BSSs in each sequence are still lacking.
\item Significant progress has recently been made in observing BSSs in young OCs in the Magellanic Clouds. However, these observations are limited to a small number of clusters, making it difficult to draw definitive quantitative conclusions. Future efforts should focus on observing massive BSSs in a larger sample of young OCs.
\item More systematic observations of BSSs across different environments are crucial for statistical analyses and to bridge the gap in our understanding between high-mass and low-mass BSSs. This particularly includes additional observations of BSSs in intermediate-age clusters.
\item New observational techniques, such as asteroseismology, may provide us crucial constraints on the origin of BSSs \cite[see][for example]{2023AJ....165..188G}, as merger products and those formed through stable mass transfer are expected to exhibit distinct internal structure and, consequently, different asteroseismic signatures \citep{2021ApJ...923..277R,2024A&A...690A..65H,2024A&A...687A.222W,2024ApJ...967L..39B}.
\end{itemize}

\begin{ack}[Acknowledgments]

CW acknowledges funding from the Netherlands Organisation for Scientific Research (NWO), as part of the Vidi research program BinWaves (project number 639.042.728, PI: de Mink). The authors thank the editor, Fabian Schneider, for his constructive feedback and valuable comments.
\end{ack}

\seealso{\cite{1993PASP..105.1081S,1995ARA&A..33..133B,2015ASSL..413.....B}}


\bibliographystyle{Harvard}
\bibliography{reference}
\end{document}